\documentclass{article}

\usepackage{PRIMEarxiv}
\usepackage{orcidlink}
\usepackage[utf8]{inputenc} 
\usepackage[T1]{fontenc}    
\usepackage{hyperref}       
\usepackage{url}            
\usepackage{booktabs}       
\usepackage{amsfonts}       
\usepackage{nicefrac}       
\usepackage{microtype}      
\usepackage{lipsum}
\usepackage{fancyhdr}       
\usepackage{graphicx}       
\graphicspath{{media/}}     
\usepackage{amsmath}
\usepackage{multirow}
\usepackage{booktabs}

\newcommand{\thinnermidrule}{\midrule[\lightrulewidth]}

\title{Automatically Generating Narrative-Style
Radiology Reports from Volumetric CT Images;
a Proof of Concept
}

\author{
  Marijn Borghouts~\orcidlink{https://orcid.org/0009-0002-3820-3957} \\
  Department of Biomedical Engineering \\
  Medical Imaging Group \\
  Technical University of Eindhoven \\
  The Netherlands \\
  \texttt{m.m.borghouts@student.tue.nl} \\
}

\begin{document}
\maketitle

\begin{abstract}
The world is facing a shortage of radiologists, leading to longer treatment times and increased stress, which can negatively affect patient safety and workforce morale. Integrating artificial intelligence to interpret radiographic images and generate descriptive reports offers a promising solution. However, there is limited research on generating natural language descriptions for volumetric medical images.
This study introduces a deep learning-based proof of concept model aimed at accurately identifying abnormalities in volumetric CT data and subsequently generating corresponding narrative-style reports. Various encoder-decoder models were assessed for their efficacy in addressing both clinically relevant and surrogate tasks. Clinically relevant tasks encompassed identifying and describing pulmonary nodules and pleural effusions, while surrogate tasks involved recognizing and describing artificial abnormalities such as mirroring, rotation, and lung lobe occlusion.
The results demonstrate a remarkably high accuracy in detecting combinations of artificial abnormalities, with the best model reaching a classification accuracy of 0.97 on an independent dataset with a homogeneously distributed 11 class problem. Furthermore, the best model consistently generated coherent radiology reports in natural language, with a next-word prediction accuracy of 0.84. On top of that, 65\% of these reports were factually accurate regarding the identified artificial abnormalities. Unfortunately, these models did not replicate this success for clinically relevant tasks.
Overall, this study manages to provide a working proof of concept model for a challenge yet to be addressed by the scientific community. Given the achieved success on the surrogate tasks, the leap to clinically relevant tasks seems within reach. The acquisition of a significantly larger high quality dataset appears to be the most promising path to making this leap. Furthermore it is expected that performance will increase significantly when more computational resources become available to train this model end-to-end.

\end{abstract}


\section{Introduction} \label{sec:introduction}
In 2022, the Royal College of Radiologists of the United Kingdom (UK) released a report focused on the status of the clinical radiology workforce in the UK \cite{rcr-census}. This report highlight a concerning 29\% deficit in radiologists, a trend that is projected to intensify in the coming years. As a result, 90\% of clinical
directors in UK cancer centers are concerned that workforce
shortages will negatively impact patient safety due to treatment delays and stress-induced errors. On top of that, 100\% of those directors are concerned about staff morale and burnout. The prevailing staff shortage places an overwhelming burden on the remaining radiologists, further increasing the risk of burnout and reducing the workforce in a downward spiral. These alarming reports are just one of many which address similar problems worldwide \cite{Konstantinidis2023, IHS2021, Farmakis2021, DeBenedectis2022}.

To address these critical issues and guarantee a sustainable radiology workforce, the integration of artificial intelligence (AI) emerges as a promising solution. The potential of AI to (partially) automate the radiographic image analysis process offers a reduction in workload for radiologists \cite{Shoshan2022, Tong2023, Sim2023}, mitigating delays in patient treatment and preventable human errors, subsequently improving patient survival rates. Moreover, automating radiographic image analysis contributes to standardizing the workflow, reducing both inter- and intra-observer variability \cite{Zhao2022}. Lastly, the resulting standardized workflows are more  suitable to optimization, leading to a reduction in costs \cite{Sim2023}. In summary, the proper implementation of artificial intelligence to (partially) automate radiographic image analysis stands to benefit both medical professionals, patients, and tax-payers.

AI for radiographic image analysis took flight after the success of AlexNet in the 2012 ImageNet Large Scale Visual Recognition Challenge (ILSVRC) \cite{Krizhevsky2012}, igniting a new era of research and development centered around deep convolutional neural networks (CNNs). Since the mid-2010s, CNNs have become the cornerstone of (radiographic) image analysis. 

Although image analysis remains the focal point in AI-driven radiology studies, it only encompasses a portion of a radiologist's responsibilities in processing radiographic images. After analysis, radiologists must often articulate their findings in comprehensive reports to facilitate communication with other medical professionals. 

These reports typically adopt either a structured or narrative-style format. Structured reports adhere to predetermined templates, often comprising closed questions with a fixed set of possible answers. Conversely, narrative-style reports are more free-form, resembling traditional written narratives, wherein radiologists convey findings in coherent natural language. Notably, narrative-style reports hold broader applicability compared to structured reports, as they accommodate all possible clinical abnormalities, something which does not hold for a standardized set of closed questions.

Research into narrative-style radiology report generation is a relatively new field. This is in part due to its dependence on natural language processing (NLP) techniques capable of generating long coherent pieces of text; techniques which are relatively new compared to the CNN techniques that are required to answer closed questions. Note that closed question answering can be performed using image classification with some simple post-processing, writing narrative-style reports cannot. The first NLP architectures capable of ``long'' sequence generation are recurrent neural networks (RNNs) such as the Long-Short-Term Memory (LSTM) model which was introduced in 1997 \cite{Hochreiter1997}. However it took 20 more years until the advent of the transformer architecture in 2017 \cite{Vaswani2017} before a significant leap forward was made in AI language generation. This new transformer architecture paved the way for state-of-the-art (SOTA) large language models (LLMs) such as the "Bidirectional Encoder Representations from Transformers" (BERT) \cite{Devlin2019} and the "Generative Pre-trained Transformer" (GPT) \cite{Radford2018}. These SOTA models enable longer sequence generation due to their efficient computations, allowing them to better attend to long input sequences

A substantial amount of research has focused on generating narrative-style reports from 2D images (see related works in Section \ref{sec:related_works}). Little research has been done on converting 3D images into structured reports. However, to the best of my knowledge, not a single peer reviewed study exists which succeeds in converting 3D images into narrative-style reports. Hence this work aims to fills that research gap by being among the first to generate narrative-style reports from volumetric CT images.

The field of narrative-style report generation from volumetric imaging data remains severely underdeveloped. Hence, the aspiration to develop a clinically viable AI-tool capable of unburdening human radiologists remains out of reach. Instead, this work serves as an exploratory study, seeking to discover what is achievable with student-level resources, and discover were the biggest challenges lie. Thus, the primary objective of this work is to develop a proof of concept that stimulates others to continue research in this young field, and to provide a framework that is user-friendly, modular, and well documented; thereby facilitating a convenient starting point for subsequent research endeavors. The envisioned proof of concept model should be able to recognize and describe abnormalities in radiographic images. Given the exploratory nature of this work, abnormalities and their narrative-style descriptions were synthetically generated to facilitate the creation of a dataset that is both larger and more balanced than real-world equivalents. This synthetic dataset comprises "artificial abnormalities," which are visually easier to discern compared to their clinically relevant counterparts. Additionally, the narrative descriptions of these artificial abnormalities employ simpler natural language than that which is typically found in clinical radiology reports. Nonetheless, efforts are also made to apply the developed approach to clinically relevant cases.

More specifically, the research goal is to \textbf{accurately detect abnormalities in volumetric CT images and to subsequently generate a narrative-style report, accurately describing these abnormalities.} The proposed research goal appears feasible, as demonstrated by 1) existing research into solving this task for 2-dimensional images \cite{Yin2019, Kaur2022, Nimalsiri2023, DallaSerra2024, DallaSerra2023, Tanida2023, Xue2024, Chen2020, Jing2017, Liao2023, Monshi2020} 2) studies which successfully employ CNNs to detect various diseases from volumetric images \cite{Dai2022, Zhou2019, Draelos2021}, and 3) the recent rise of LLMs, such as OpenAI's ChatGPT, which highlight the capability of AI to produce substantial pieces of coherent natural language conditioned on specific inputs \cite{Ray2023}.

\section{Technical Background} \label{sec:background}
    \begin{figure*}
    \centering
    \includegraphics[width=\textwidth]{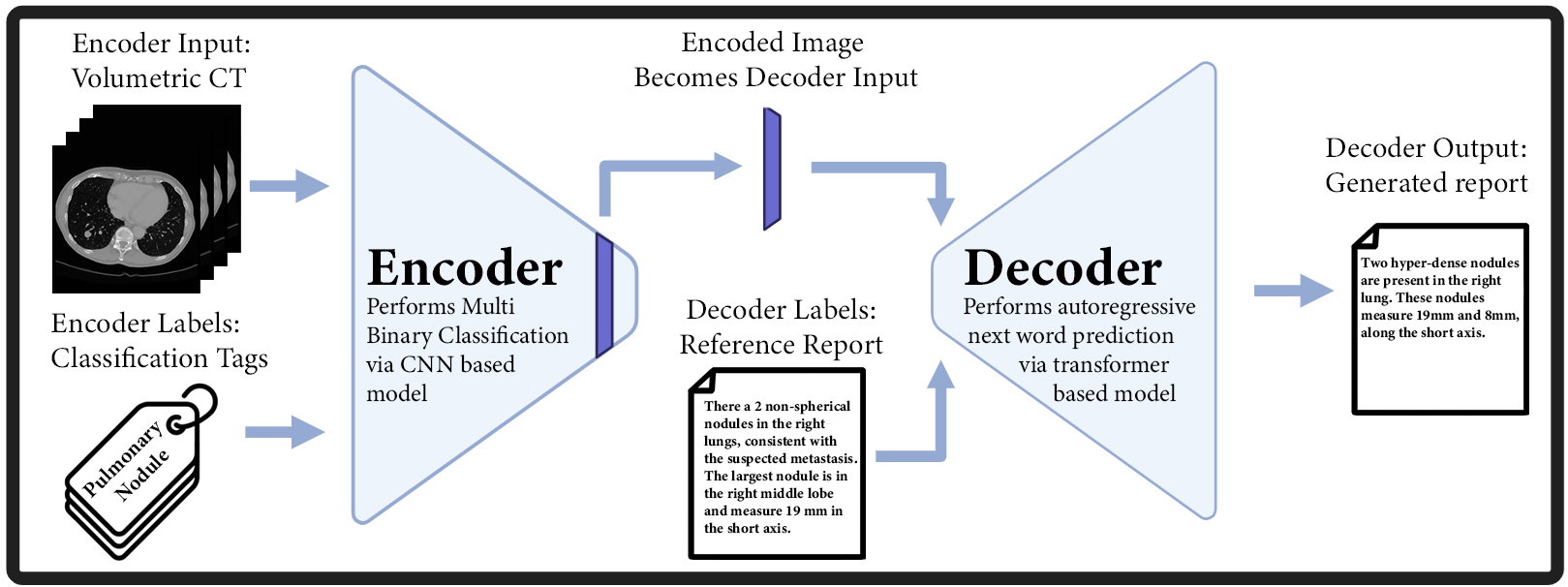}
    \caption{Schematic overview of the proposed encoder-decoder approach.}
    \label{fig:architecture}
\end{figure*}

Automatic radiology report generation (ARRG) is, in essence, an advanced form of automatic image captioning. The modern, deep learning-based, field of automatic image captioning gained significant momentum with the 2015 publication titled "Show and Tell: A Neural Image Caption Generator" \cite{Vinyals2015}. In this pioneering approach, a CNN encodes images into a lower-dimensional latent space, which a LSTM model then decodes into natural language.

CNNs are particularly well-suited for image compression due to their convolutional layers, where a kernel or so-called filter slides over the image, performing element-wise multiplications and averaging to produce an output that is smaller than the input. This operation reduces image size while preserving essential features, making it highly effective for image encoding. By stacking multiple convolutional and pooling layers, CNN models can compress images into a few output values. These value usually correspond with the different classes in case the CNN is trained as an image classifier. Properly trained CNNs can learn to classify images of almost everything, given that the object in question can be seen by eye \cite{Russakovsky2015}. Early convolutional layers of the CNN detect simple features like edges or geometric patterns, while deeper filters combine features from previous layers to recognize more complex entities \cite{Zeller2014} such as pulmonary tumors in CT images. CNNs' use of shared weights for image filters enhances their computational efficiency compared to Multi-Layer Perceptrons (MLPs), and the local sharing of filters improves their ability to learn generalizable features \cite{Goodfellow2016}. CNNs are computationally efficient due to their use of shared weights in the filters \cite{Goodfellow2016}. Furthermore, filter are applied locally, which improves their ability to learn generalizable features as compared to overfitting to global overly complex features \cite{Goodfellow2016}. These are two of the main attributes which makes CNNs superior to Multi-Layer Perceptrons (MLPs) for image encoding.

In the decoder, the generation of image captions occurs autoregressively when using a RNN or transformer-based architecture. Autoregressive generation means that the caption is created word by word in a sequential manner. For instance, at t=0, the decoder accesses the encoded image and the start-of-sequence (SOS) token, outputting a probability distribution over all possible tokens in its vocabulary. Tokens typically represent words or sub-parts of words. The ground truth at t=0 is also a probability distribution, with zero probabilities for all tokens except the first token of the reference report, which has a probability of one. The model is trained using binary cross-entropy loss (BCELoss), as shown in Equation \ref{eq:binary_cross_entropy}.

\begin{equation}
\mathcal{L} = -\frac{1}{N} \sum_{i=1}^{N} \left[ y_i \log(p_i) + (1 - y_i) \log(1 - p_i) \right]
\label{eq:binary_cross_entropy}
\end{equation}

Where $\mathcal{L}$ is the binary cross-entropy loss and N is the number of tokens in the vocabulary. $y_i$ denoted the true probability for the ith token (0 or 1) and $p_i$ represents the predicted probability for the ith token (float between 0 and 1). This equation calculates the average cross-entropy loss over all N tokens in the vocabulary. The model is optimized to minimize the BCELoss, effectively minimising the difference between the ground truth and predicted probability distribution. At t=0, the predicted token is the token with the highest probability. Next the cycle will repeat to generate the second report token. At t=1, the decoder has access to the encoded image, the SOS token, and the token predicted at t=0. At t=2, the input for the decoder consists of the encoded image and the tokens generated at t=0and t=1. This sequence continues until the end-of-sequence (EOS) token is generated or until a predetermined number of steps is reached.

While the "Show and Tell" paper \cite{Vinyals2015} employed an RNN decoder, modern approaches favor transformers architectures. The power of the transformer architecture \cite{Vaswani2017} lies in its attention mechanism. Specifically, the transformer block uses a self-attention layer to learn contextual relationships between words, meaning it learns which words often appear in the same sentences and in which order or structure they appear. For instance, when trained correctly, a transformer model can discern from the context, i.e. the surrounding word, whether the word "bank" refers to a financial institution or a mount of sand near a river. Additionally, by embedding the position of a word within its sentence into the vector that represents the word, the transformer can process entire sequences in parallel, unlike RNNs which process sequences sequentially. These two major advantages make transformers, compared to RNNs, a more powerful architecture for generating coherent text while attending to a given input.

Most of the research in the (radiographic) image captioning field stuck to this encoder-decoder approach \cite{Liao2023}, and this study is no exception. However, this study will pioneer the translation of volumetric images into narrative-style reports. Training such deep learning models presents several challenges, with the primary challenge being the acquisition of a sufficiently large dataset, which can be difficult due to the sensitive nature of medical data and the associated privacy laws. Additionally, managing the substantial memory requirements of volumetric data and the associated computational power necessary for processing such data poses a significant challenge. A volumetric CT image is significantly larger than its 2D counterpart, often by hundreds of times. To capture the rich information in the 3D volume using a CNN, a set of (pseudo) 3D convolutional kernels are required. Again, the 3D variant of a kernel is substantially larger than its 2D counterpart. As a result, a high-end personal graphics card with 12GB of video RAM, costing hundreds of dollars, is insufficient to initialize even a relatively shallow 3D CNN, LLM, and a single volumetric image. This poses a considerable challenge to this work which has to rely on one such graphics card. Unless numerous GPUs are available for distributed training, alternative strategies must be implemented to address this challenge.
 
This work proposes training the encoder and decoder separately, allowing for more manageable memory requirements and easier performance evaluation of each model stage. Figure \ref{fig:architecture} provides an overview of the proposed methodology. Section \ref{sec:methods} "Methods" provides a detailed methodology for the proposed approach.

\section{Related Works} \label{sec:related_works}
In recent years, the field of radiology has seen significant advancements in automatic radiology report generation systems. A substantial amount of research has focused on addressing this challenge in the realm of 2D images \cite{Yin2019, Kaur2022, Nimalsiri2023, DallaSerra2023, DallaSerra2024, Tanida2023, Xue2024, Chen2020, Jing2017}, particularly chest X-rays (CXR), with systematic review studies \cite{Liao2023, Monshi2020} documenting these efforts. 

The current SOTA research in the two dimensional CXR domain has surpassed the simple encoder-decoder structure, employing more sophisticated approaches to enhance performance. For instance, the 2023 work of \cite{Tanida2023} implemented an object detection model to identify 29 anatomical regions in CXRs. They then used binary classifiers to select the most interesting regions. The encoded representations of these regions were utilized to generate report sentences. This method not only improved the accuracy of report generation but also enhanced the explainability of AI systems by allowing each report sentence to be traced back to a specific anatomical region. 

Similarly, the 2024 work of \cite{Xue2024} leveraged auxiliary signals derived from the most frequently occurring disease-related words in the training reports to amplify features associated with abnormal regions in the images. This approach aimed to mitigate data bias towards healthy regions. Additionally, they incorporated a memory mechanism-driven decoding module, which reintroduced the auxiliary signals into the decoding process to improve the contextual consistency of the generated reports.

The examples above demonstrate the significant technical advancements in the 2D report generation domain. However, this study aims to investigate the feasibility of employing an encoder-decoder methodology for translating \textbf{volumetric} imaging data into \textbf{narrative-style} reports. To the best of my knowledge, only two studies have addressed this challenge.

The first study, by \cite{Jiang2018}, utilized a CNN-LSTM architecture to describe liver tumors from volumetric CT images. Although this study technically processed 3D images, it treated the 2D slices of the 3D images individually. By employing an attention weighting mechanism, the model aggregated features from the individual slices, focusing on the most relevant ones. The LSTM decoder was then conditioned on these attention-weighted features. The generated reports, while technically narrative-style, followed a structured format describing the tumor's shape, contour, and intensity, in that order.

The second study, an 2024 ArXiv pre-print by \cite{Hamamci2024}, introduced the CT2Rep model, which generates true free-form narrative-style reports from volumetric CT chest images, describing various findings. The CT2Rep model utilizes a custom transformer-based encoder that divides the 3D volume into smaller 3D patches, generating latent space representations for each while preserving the three-dimensional relationships. The decoder, also transformer-based, includes a "relational memory" module to capture and propagate information patterns throughout the decoding process.

\section{Methodology} \label{sec:methods}
    \subsection{Dataset}
Obtaining a sufficiently large dataset of volumetric CT data with both classification labels for the encoder's classification task and narrative-style reports for the decoder's text generation task is essential, given the data-intensive nature of deep learning. Publicly available medical datasets are limited due to the sensitive nature of medical data, necessitating strict data anonymization procedures or restricted access. Unfortunately, publicly available volumetric CT datasets featuring this dual-label configuration are scarce, if not non-existent. Consequently, the dataset for this work had to be sourced from an institution willing to share their private data.

This work used a dataset obtained from the imaging department of the University Medical Center Utrecht (UMCU). Although provided by the UMCU, the scans and reports were generated by eight different Dutch medical centers (Amphia, Isala, Leiden University Medical Center, Maxima, Radboud, Groningen University Medical Center, Amsterdam University Medical Center, Zuyderland), involving diverse medical staff, equipment, protocols, and report templates. The dataset consisted of unprocessed images and radiology reports collected from patients who had previously been diagnosed with melanoma and underwent medical imaging to detect potential metastasis. The primary imaging modality used for metastasis detection was low dose CT scans (LDCT), often in conjunction with positron emission tomography (PET) scans. LDCT scans can reveal indications of metastasized cancer, such as enlarged lymph nodes or hyper- and hypo-dense nodules. Additionally, LDCT provides contextual information for PET scans, which highlight metabolically active tissues typically associated with growing tumors. These modalities can detect different types of true positives, thereby working synergistically \cite{Ye2018}. The dataset exclusively comprises LDCT images. The reports were either LDCT reports or LDCT-PET reports. 

The raw images underwent pre-processing to render them suitable for input in the encoder model. In this study, it was decided to exclusively focus on images of the thorax, partly due to the thorax being the most frequently imaged body region in the dataset and partly because restricting the analysis to the thorax reduces both image size and the potential number of abnormalities under consideration. During pre-processing, the images were standardized in orientation. Subsequently, the lung were segmented using the TotalSegmentor \cite{Wasserthal2023} model. The collection of transverse slices which made up the thorax, was defined as 100\% of transverse slices which made up the lung segmentation, additionally,  40\% of this slice count was selected both above and below the lungs. This approach ensures that the thorax size is relative to the lung size, enhancing robustness against variations in imaging settings, image resolution, and body sizes. The thorax thus consists of 40\% + 100\% + 40\% = 180\% of the slices of those in the lung segmentation. Following thorax segmentation, the images were rescaled to optimally fit the image dimensions required by the encoder model, while preserving aspect ratios. Given that the original images may not possess the same aspect ratio as the required image size, padding was applied in the necessary dimensions until the desired image size was achieved. Subsequently, pixel values of the images were clipped to range [-1000, 1300] in Hounsfield units (HU) and then normalized. Finally, pixel values were centered on the ImageNet mean, in case the encoder model was initialized using ImageNet weights. After pre-processing, 464 usable images remained to form the dataset of which 364 were used for training, 50 were used for validation, and 50 were used for the final independent test set.

\begin{figure*}
    \centering
    \includegraphics[width=\textwidth]{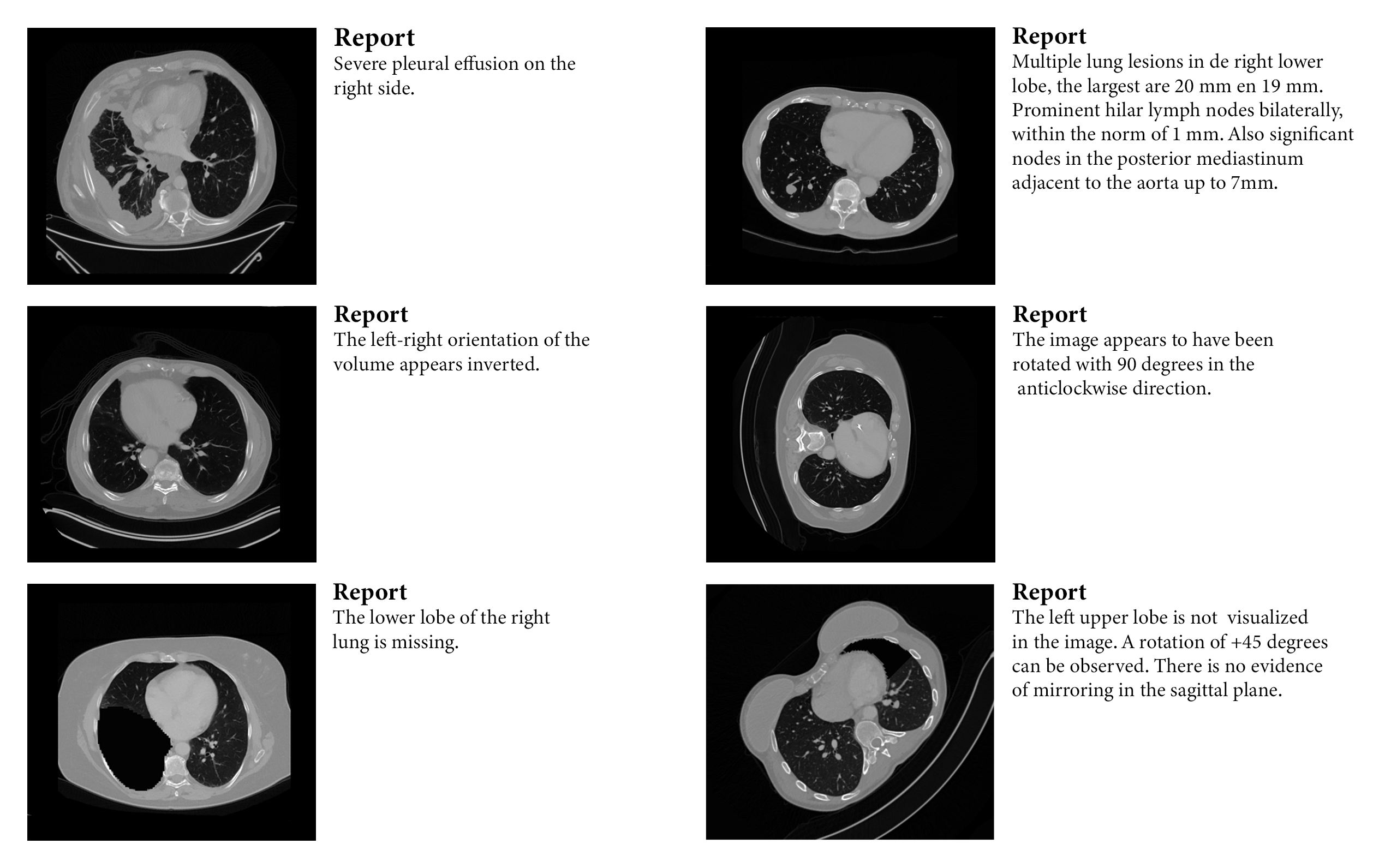}
    \caption{Dataset examples: Transverse slices of volumetric CT images and their corresponding narrative-style report. One for investigated task. Top left: pleural effusion. Top right: pulmonary nodule. Middle left: mirroring. Middle right: rotation. Bottom left: lobe occlusion. Bottom right: Combined surrogate tasks.}
    \label{fig:input_data_example}
\end{figure*}

The raw reports, which were used for the clinically relevant tasks, underwent pre-processing to render them usable within the decoder model. Due to the diverse origins of these reports, originating from various hospitals with differing formats and authored by different medical personnel, automating the report pre-processing proved extremely challenging. Consequently, manual pre-processing was conducted. Numerous pieces of information were removed from the reports during manual pre-processing. This included the removal of personal information regarding the patient and radiologist. Information regarding the scanner type, imaging protocols, pre-imaging checks, and examination dates was also removed. Medical history was retained only if it was currently observable in the CT scan. References to prior medical examinations were also removed. Findings related to PET scans were omitted, as were any findings outside the thoracic region (which was defined as ranging between the thyroid and the diaphragm). Additionally, references to specific image slices were eliminated, since this correspondence was lost due to prior rescaling and padding performed during image pre-processing. Figure \ref{fig:input_data_example} shows one example image and the corresponding reference report for every task which was investigated in this study.

\subsection{Surrogate Tasks}
This study aims to establish a proof of concept within the constraints of a limited dataset comprising fewer than 500 images. Surrogate tasks share characteristics with the primary task but are easier to solve. By training the model on these surrogate tasks the likelihood of generating a working proof of concept is greatly enhanced. Furthermore, by achieving good performance on the surrogate tasks it can be more convincingly argued that the primary task is within reach. Synthetically generated "artificial abnormalities" are used as surrogate tasks in this study because they are easier to recognize and describe compared to clinically relevant abnormalities.
By nature of being artificially created, the artificial abnormalities exhibit greater visual consistency across samples and they are evenly distributed to achieve a balanced dataset. 

The artificial abnormalities selected for this study were "mirroredness", "rotation", and "lung lobe occlussion". Mirroredness involved flipping the image in the sagittal plane, making the left side of the patient appear as the right side and vice versa. This transformation was randomly applied to 50\% of the images and treated as a single-class classification task. Rotation entailed either a -90, -45, 0, 45, or 90 degree rotation of the image. Rotation was treated a classification task with five classes. Each image was randomly rotated by one of the five classes ensuring equal class distribution. Lung lobe occlusion involved the complete masking of one of the five lung lobes. Lobe occlusion was treated as a classification task with five classes. Each image was randomly treated with one lobe mask, ensuring equal class distribution. The lung lobe mask used for this process were generated by the TotalSegmentor model \cite{Wasserthal2023}. In case of lobe occlusion and rotation, five different images could be created from each original image. Hence, for these tasks the dataset size was expanded to 464*5=2320 images, offering a significant increase in dataset size.

The narrative-style reports for the individual artificial abnormalities consisted of a single sentence describing each respective abnormality. For each abnormality, over 30 template sentences were available. For example, a template sentence for the rotation task could be: \textit{"The image appears to have been affected by a [direction] rotation of [degrees] degrees."} Based on the specific class, the template sentences would be filled in to become, for instance: \textit{"The image appears to have been affected by a counterclockwise rotation of 45 degrees."}

The three individual surrogate tasks were used to validate which models and hyperparameter variations performed best. Once the best-performing model was identified, it was tested one last time on an independent unseen dataset, where all three surrogate tasks were combined in a single image. This meant that each image could simultaneously exhibit mirroring, rotation, and lobe occlusion. The final combined surrogate task included 11 simultaneous binary classification tasks, with findings described in reports of three sentences and up to 50 words in total. This combined surrogate task closely approximates the primary task and, therefore, appears to be a reasonable proxy for possible success in the main task.

\subsection{Label Mining}
To train the encoder effectively, a set of classification labels was required. These labels were not readily available for the clinically relevant tasks. However, information about the presence or absence of specific abnormalities was embedded in the narrative-style reports. In this study, these classification labels were automatically extracted from the reports.

To accomplish this task, an adapted version of the SARLE (Sentence Analysis for Radiology Label Extraction) tool was employed \cite{Draelos2021}. SARLE is a conceptually straightforward, rule-based algorithm designed for mining classification labels from narrative-style reports.

The SARLE algorithm consists of two main phases. The first phase involves removing all (sub)sentences that describe medically normal findings. This is accomplished through a rule-based system, wherein the algorithm iterates over all words in search of designated "main words." Each main word is associated with a predefined operation. For instance, when encountering the main word "no sign of," the algorithm deletes all words following the main word until it reaches a full stop or the word "however." In the second phase, the SARLE algorithm examines the remaining sentences for abnormalities using a medical vocabulary. For a more detailed description of SARLE's methodology, see the work of \cite{Draelos2021} or the code at \href{https://github.com/Marijn311/CT-Report-Generation}{https://github.com/Marijn311/CT-Report-Generation}. 

One of the major advantages of SARLE's rule-based approach is its ability to function without manually labeled data. However, SARLE's functionality can be expanded in scenarios where at least a part of the reports are manually labeled. If given a training set containing labels indicating whether a sentence is medically normal or abnormal, the first phase of SARLE can be executed by training a machine learning classifier, thereby replacing the rule-based approach. Furthermore, if a validation set with ground truth classification labels is provided, the model's label extraction accuracy can be assessed. Another notable advantage of the conceptually simple, rule-based approach is its flexibility, allowing users to easily customize the medical vocabulary and associated rules, thereby enabling adaptation for mining any desired label.

The utilized dataset comprises volumetric CT scans sourced from patients previously diagnosed with melanoma. Consequently, the dataset predominantly contains abnormalities associated with metastatic cancer. Based on their prevalence and visual discernibility, two clinically relevant abnormalities were selected: pulmonary nodules and pleural effusion.
The dataset totaled 464 images, with 285 images labeled as "pulmonary nodule" and 37 images labeled as "pleural effusion" by the SARLE algorithm. An evaluation of SARLE's label extraction performance is given in Appendices \ref{appendices:SARLE}.
    
\subsection{Model Architectures}
The main research objective encompasses two primary components: 1) the recognition of (artificial) abnormalities within CT images, and 2) the generation of narrative-style reports detailing these abnormalities. The architectural framework adopted for this purpose is structured accordingly. The proposed approach adopts an encoder-decoder paradigm, wherein abnormality recognition is executed by the encoder, and report generation is facilitated by the decoder. The encoder segment employs a CNN feature extractor, extended by a classifier, to address a multi-label classification task. This phase, fittingly termed the encoder, aims to capture the three-dimensional image data into a lower dimensional latent space. The decoder segment employs a transformer-based model, tasked with generating a radiology report. The report generation in the decoder is conditioned on the latent space representation of the corresponding image and on the previously generated words of the report. See Figure \ref{fig:architecture} for a schematic representation of the proposed approach. This study investigates the performance of two CNN-based feature extractors, three classifiers, and three transformer-based decoders. The following paragraphs very briefly describe the main idea of every architecture. For more details on the exact model architectures, please see the implementation on the Github page for this work: \href{https://github.com/Marijn311/CT-Report-Generation}{https://github.com/Marijn311/CT-Report-Generation}, or see the references to the original works.

The first feature extractor, named CT-Net, is based on the work of \cite{Draelos2021}. CT-Net splits the volumetric gray-scale image into chunks of three transverse slices. Each chunk of three slices is passed to a 2D RGB ResNet18 that has been initialized with ImageNet weights. CT-Net therefore produces a separate set of feature maps for every chunk of three transverse image slices. 

The second feature extractor, named Medical-Net, is based on the work of \cite{Chen2019}. Medical-Net  utilizes a fully 3D ResNet18 that has been pre-trained on a multi modal medical imaging dataset. Medical-Net takes an entire gray-scale volumetric image and extracts one set of feature maps per image. 

The first classifier, named 3D-Convs, utilizes three consecutive 3D convolution layers followed by three linear layers to shape the output into the desired shape. This classifier is based on the one used the CT-Net work \cite{Draelos2021}.

The second classifier, named Attention Pooling, flattens feature maps into vectors and applies an attention mechanism to generate attention scores. These scores are used for attention-weighted average pooling, before the average feature vector is passed through a fully connected layer to shape the output into the desired shape.

The third classifier, named Transformer, employs a transformer block with self-attention to compute similarity scores between feature vectors. These scores facilitate the updating of feature vectors to capture context. These steps are followed by attention-weighted pooling and a fully connected layer akin to the Attention Pooling Classifier. 

The first decoder, named Transformer Scratch, comprises six stacked transformer blocks, totaling 4M trainable parameters. It integrates cross-attention modules within each block, conditioning report generation on latent space image representations. The model was initialized with random weights.

The second decoder, named Bio+Clinical BERT, underwent a two-step pre-training: first on PubMed data, resulting in BioBert, then on the MIMIC dataset which contains radiology reports, for more details see \cite{Alsentzer2019}. The BERT model is loaded as a decoder by adding cross-attention modules to condition report generation on latent space image representations. Bio+clinical BERT consists of 12 stacked transformer blocks, totalling 159M trainable parameters.

The third decoder, named BioGPT, was pre-trained on PubMed abstracts, for more details see \cite{Luo2022}. This decoder does not integrate cross-attention modules. Instead, the encoded images, in token representation, are added as the first part of the reference report. Predicted image tokens are removed from the model's output before being passed to the loss function, thereby ensuring that the model is neither trained nor penalized for generating image tokens. The BioGPT model is the largest of the three decoder models, having 24 stacked transformer blocks and 346M trainable parameters.

\subsection{Experiments}
Processing volumetric imaging data through a CNN requires significant computational power due to the larger size of volumetric CT images and the increased parameters of 3D kernel models. Due to limited GPU resources, the decision was made to freeze the weights of the CNN models, and only update classifier parameters, leveraging pre-trained weights of the feature extractors.
  
The experiments are divided into a distinct encoder and decoder segment. For the encoder component, the investigation encompasses two feature extractors and three classifiers, resulting in six potential combinations. However, the Attention Pooling classifier and the Transformer classifier require sets of feature vectors to operate. While the CT-Net feature extractor yields a feature vector for each segment composed of three consecutive image slices, the Medical-Net model generates a single feature vector per image. This means that the Medical-Net feature are incompatible with those two classifiers. As a result only four, as opposed to six, distinct encoder models were evaluated for their efficacy in detecting anomalies in CT images. The encoder models were trained on a variety of classification tasks, encompassing two clinically significant challenges (detecting pleural effusion and pulmonary nodules) and three surrogate tasks (detecting mirroredness, rotation, and lobe occlusion).

The training dynamics and resulting model performance can be greatly influenced by the hyperparameter configuration. To keep the amount of training runs in this study manageable it was decided to only vary the batch size and the learning rate. ADAM was used as the optimizer and no learning rate scheduler was applied. It was decided to vary the learning rate between 1e-2 and 1e-5. The batch size was varied between 6 and 50. First, the middle of this hyperparameter grid was explored. Based on the training dynamics, the hyperparameters were adjusted and the models were retrained until either the validation accuracy reached at least 0.95 or 4 different hyperparameter combinations were explored without any of the training runs being able to decrease the validation loss. All encoder models were trained with at least 1500 parameter update steps or until the validation loss started increasing.

The performance in classifying clinically relevant abnormalities was evaluated using the area under the precision-recall curve. This metric is particularly advantageous for imbalanced datasets, a common occurrence in medical image datasets with abnormality labels. The precision-recall curve shows the classification performance related to positive samples, making it well-suited for datasets where the positive class is infrequent. Precision (Equation \ref{eq:precision}) signifies the ratio of true positive predictions to all positive predictions, while recall (Equation \ref{eq:recall}) denotes the ratio of true positive predictions to all actual positives. Notably, the area under the precision-recall curve remains robust against variations in the chosen threshold for binarization. Conversely, the classification of artificial abnormalities was assessed using accuracy. Accuracy (Equation \ref{eq:accuracy}), being an intuitive metric, indicates the proportion of correct predictions out of all predictions made. Accuracy may yield misleading results for imbalanced datasets, however, the prevalence of artificial abnormalities is uniformly distributed, mitigating this limitation.

\begin{equation}
\text{Precision} = \frac{TP}{TP + FP}
\label{eq:precision}
\end{equation}

\begin{equation}
\text{Recall} = \frac{TP}{TP + FN}
\label{eq:recall}
\end{equation}

\begin{equation}
\text{Accuracy} = \frac{TP + TN}{TP + FP + FN + TN}
\label{eq:accuracy}
\end{equation}

Due to time constraints is was not feasible to evaluate all encoder-decoder combination for all tasks. Therefore, based on the results of the encoder experiments (Section \ref{sec:results}, Table \ref{tab:surrogate_encoder_accuracy}) it was determined that the CT-Net + 3D Convs encodings would be used as input for all decoder models. The three previously defined decoder models were trained on the individual surrogate tasks. The models were not trained on the clinically relevant tasks since the encoder models failed to accurately encode the images for these tasks. Based on the results for these decoder experiments (Section \ref{sec:results}, Table \ref{tab:surrogate_decoder_accuracy}, and Table \ref{tab:surrogate_decoder_factual_accuracy}) the most promising decoder model underwent additional training to describe a combination of the three surrogate tasks on an independent test set.

The performance of the decoder models is evaluated using the accuracy of the next-word prediction, calculated as the ratio of correct predictions to the total number of words in the generated report. Given the inherent freedom and variability of natural language, coherent and correct reports may be generated that differ significantly from the ground truth. Hence, the decision was made to evaluate next-word prediction accuracy whilst teacher forcing was active. On the other hand, generated reports may closely match the ground truth whilst incorrectly attending to the encoded images. For example, the prediction \textit{"The image has been rotated by 90 degrees."} closely matches the ground truth \textit{"The image has been rotated by -45 degrees."} This prediction will result in a very high accuracy, but it has incorrectly attended to the image and contains false information. Hence, to address this, a factual accuracy metric was introduced. This metric, evaluated manually using 25 randomly selected predictions, measures the proportion of factually correct predictions out of the total. 

Similar to the encoder models, it was decided to only vary the batch size and the learning rate. This variation was done within the same range as mentioned for the encoder models. Again, ADAM was used as the optimizer and no learning rate scheduler was applied. Based on the training dynamics, the hyperparameters were adjusted and the models were retrained until either the next word prediction accuracy reached at least 0.95 or 4 different hyperparameter combinations were explored. All decoder models were trained with at least 4000 parameter update steps or until the validation loss started to increase.

\section{Results} \label{sec:results}
The classification accuracy of the encoder models on the different surrogate classification tasks is shown in Table \ref{tab:surrogate_encoder_accuracy}. The combination of CT-Net + 3D Convs model generally shows the best performance across the different surrogate tasks, reaching an accuracy of 0.96 for the mirroredness task and reaching near perfect performance for the rotation and lobe occlusion tasks, with an accuracy of 1.00 and 0.99 respectively. The CT-Net + Attention Pooling model performs relatively poorly on the mirroredness and lobe occlusion task, reaching a validation accuracy of 0.81 and 0.61 respectively. However, this CT-Net + Attention Pooling combination reaches perfect classification on the rotation task. The Medical-Net + 3D Convs model, reaches an accuracy on the mirroredness and rotation task of 0.90 and 1.00 respectively. However this model was not able to learn any generalizable features for the lobe occlusion task, reporting an accuracy of 0.20, which equal random guessing over the five lung lobes. Similar behaviour can be seen for the CT-Net + Transformer model. This model reaches a validation accuracy of 0.93 on the rotation task but was unable to improve upon random guessing for the mirroredness and lobe occlusion task. Based on these results the CT-Net + 3D Convs model was determined to be the most promising encoder model. Hence this model was selected to undergo additional training on a combination of the three surrogate tasks, using an independent test set. This yielded an accuracy of 0.97. 

\begin{table*}
\centering
\caption{Classification Accuracy for Different Encoder Models on Different Surrogate Tasks. (Validation Set.)}
\label{tab:surrogate_encoder_accuracy}
\begin{tabular}{cccccc}
\multicolumn{2}{c}{ } & \multicolumn{4}{c}{\textbf{Encoder Models}} \\
\cmidrule{3-6}
\multicolumn{2}{c}{ } & \textbf{Med-Net +} & \textbf{CT-Net +} & \textbf{CT-Net +} & \textbf{CT-Net +} \\
\multicolumn{2}{c}{ } & \textbf{3D Convs} & \textbf{3D Convs} & \textbf{Attention Pooling} & \textbf{Transformer} \\
\midrule
\multirow{4}{*}{\textbf{Task}} & \textbf{Mirroredness} & \textbf{0.98} & 0.96 & 0.81 & 0.50 \\
 & \textbf{Rotation} & 1.00 & \textbf{1.00} & 1.00 & 0.93 \\
 & \textbf{Lobe Occlusion} & 0.20 & \textbf{0.99} & 0.61 & 0.20\\
 & \textbf{Combined} & N.A & 0.97\textsuperscript{*} & N.A. & N.A. \\
\bottomrule
\addlinespace
\multicolumn{6}{l}{ \textit{*Evaluated on independent test set.}} \\
\end{tabular}
\end{table*}

Secondly, the results show that not a single encoder model was able to decrease the validation loss for the pulmonary nodule or pleural effusion task. Therefore, these experiments were deemed a failure. Consequently, no metrics are reported.  

The next-word prediction accuracy of the different decoder models on the different surrogate tasks is shown in Table \ref{tab:surrogate_decoder_accuracy}. All three decoder models shows next-word prediction accuracies between 80\% and 90\% on all three surrogate tasks. The Bio+Clinical BERT models achieves the highest accuracy on all three tasks with all accuracies being 0.85 or higher. The Transformer Scratch model and the BioGPT model perform slightly worse than Bio+Clinical BERT model but they perform similar to each other. Based on the factual accuracy results in Table \ref{tab:surrogate_decoder_factual_accuracy} the Transformer Scratch model was determined to be the most promising decoder model. Hence this model was selected to undergo additional training on a combination of the three surrogate tasks, using an independent test set. This yielded an next-word prediction accuracy of 0.84 when using the feature map representation, see Appendices \ref{appendices:representation} for more details.

\begin{figure*}
    \centering
    \includegraphics[width=\textwidth]{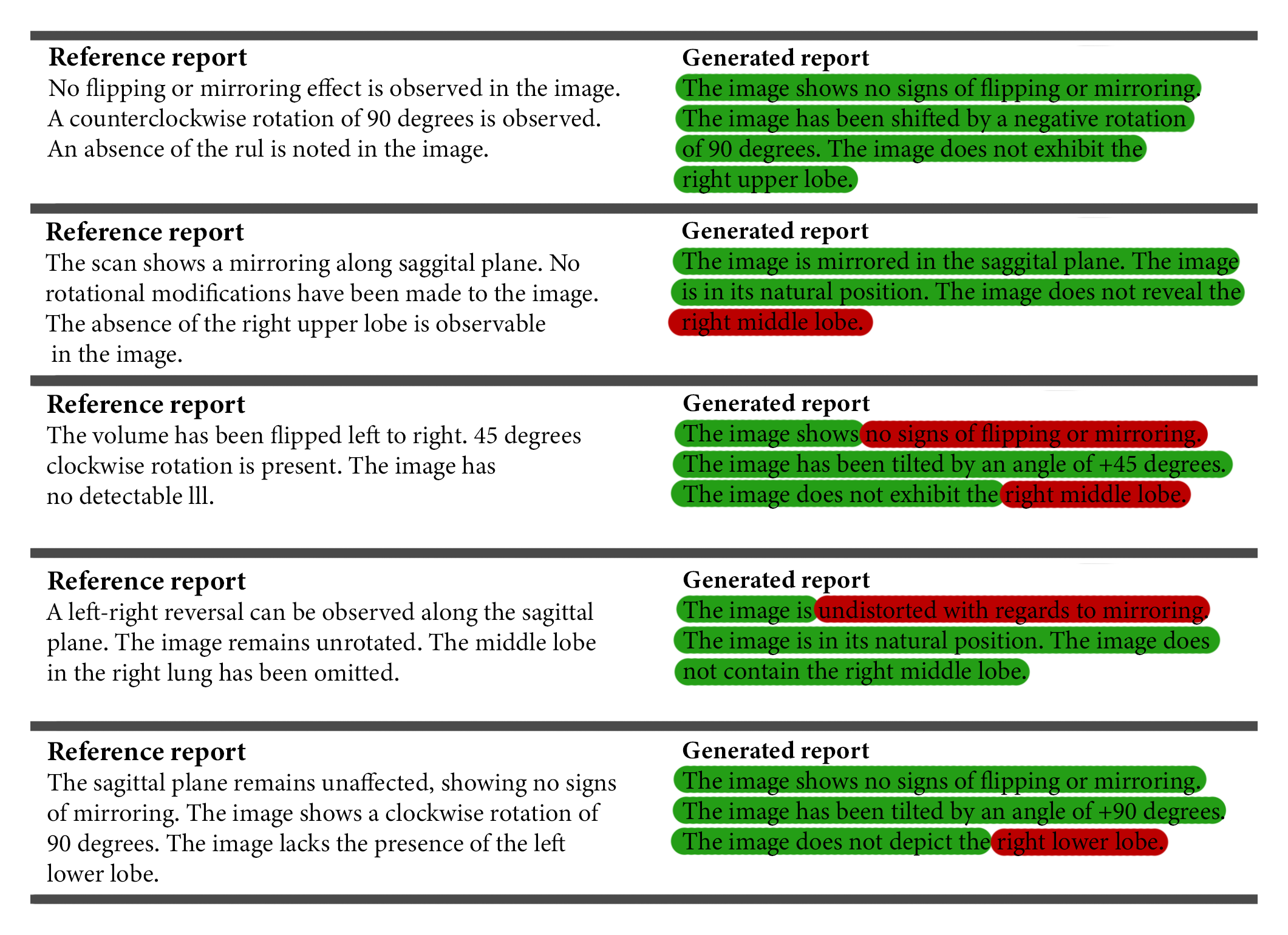}
    \caption{Five random examples of the reference report and the corresponding generated report. Generated by the CT-Net + 3D Convs + Transformer Scratch model, trained on the combined surrogate dataset. Factually accurate sentences are highlighted in green whilst inaccurate sentences are marked in red. (Examples from independent test set.)}
    \label{fig:output_data_examples}
\end{figure*}

Table \ref{tab:surrogate_decoder_factual_accuracy} shows the factual accuracy for the decoder models on the surrogate tasks. Here the results show much more variations than in Table \ref{tab:surrogate_decoder_accuracy}, again highlighting the need for metrics like factual accuracy besides word-replication metrics. The results show that the Transformer Scratch model achieves the highest factual accuracy on all three tasks. The lowest factual accuracy being 0.76 for the rotation task, followed by a factual accuracy of 0.88 for the mirroredness task, and achieving a perfect score of 1.00 for the lobe occlusion task. The performance of the Bio+Clinical BERT decoder is notably less in terms of factual accuracy, dropping to 0.60 for the rotation task and 0.68 for the mirroredness task. The BioGPT model seems to have the poorest performance with two out of three tasks having a factual accuracy under 0.50. Based on these factual accuracy results the Transformer Scratch model was determined to be the most promising decoder model. Hence this model was selected to undergo additional training on a combination of the three surrogate tasks, using an independent test set. This yielded a factual accuracy of 0.65 when using the feature map representation, see Appendices \ref{appendices:representation} for more details. Figure \ref{fig:output_data_examples} shows five random examples of the reference report and the corresponding generated report. These examples were generated by the CT-Net + 3D Convs + Transformer Scratch model, trained on the combined surrogate dataset. (Examples from independent test set.)

Since the encoder models for the clinically relevant abnormalities failed, one can assume that the encoder models did not manage to encode the images whilst retaining the relevant image content. With improperly encoded images the decoders would be conditioned on nonsense, hence no decoder models where trained for the clinically relevant abnormalities.

\begin{table*}
\centering
\caption{Next-Word Prediction Accuracy for Different Decoder Models on Surrogate Tasks. (Validation Set, Using Teacher Forcing.)}
\label{tab:surrogate_decoder_accuracy}
\begin{tabular}{ccccc}
\multicolumn{2}{c}{ } & \multicolumn{3}{c}{\textbf{Decoder Models}} \\
\cmidrule{3-5}
\multicolumn{2}{c}{ } & \textbf{Transformer Scratch} & \textbf{Bio+Clinical BERT} & \textbf{BioGPT} \\
\midrule
\multirow{4}{*}{\textbf{Task}} & \textbf{Mirroredness} & 0.88 & \textbf{0.89} & 0.86 \\
& \textbf{Rotation} & 0.82 & \textbf{0.85} & 0.80\\
& \textbf{Lobe Occlusion} & 0.83 & \textbf{0.85} & 0.84\\
& \textbf{Combined} & 0.81 / 0.84\textsuperscript{*} & N.A. & N.A. \\
\bottomrule
\addlinespace
\multicolumn{5}{l}{ \textit{*Evaluated on independent test set.}} \\ 
\multicolumn{5}{l}{ \textit{ Token / feature map representation. See Appendices \ref{appendices:representation}.}} \\
\end{tabular}
\end{table*}

\begin{table*}
\centering
\caption{Factual Accuracy for Different Decoder Models on Different Surrogate Tasks (Validation Set, Autoregressive Generation)}
\label{tab:surrogate_decoder_factual_accuracy}
\begin{tabular}{ccccc}
\multicolumn{2}{c}{ } & \multicolumn{3}{c}{\textbf{Decoder Models}} \\
\cmidrule{3-5}
\multicolumn{2}{c}{ } & \textbf{Transformer Scratch} & \textbf{Bio+Clinical BERT} & \textbf{BioGPT} \\
\midrule
\multirow{3}{*}{\textbf{Task}} & \textbf{Mirroredness} & \textbf{0.88} & 0.68 & 0.44 \\
& \textbf{Rotation} & \textbf{0.76} & 0.60 & 0.48 \\
& \textbf{Lobe Occlusion} & \textbf{1.00} & 0.84 & 0.88 \\
& \textbf{Combined} & 0.46 / 0.65\textsuperscript{*} & N.A. & N.A. \\
\bottomrule
\addlinespace
\multicolumn{5}{l}{ \textit{*Evaluated on independent test set}} \\ 
\multicolumn{5}{l}{ \textit{ Token / feature map representation. See Appendices \ref{appendices:representation}.}} \\
\end{tabular}
\end{table*}

\section{Conclusion and Discussion} \label{sec:discussion_and_conclusion}
The primary objective of this study was to develop and evaluate a proof of concept for a machine learning model capable of automatically recognizing abnormalities in volumetric CT images and describing them using natural language. Various encoder-decoder models were trained for this purpose, encompassing both surrogate tasks and clinically relevant tasks.

The results show promising performance in accurately classifying CT images with synthetic abnormalities such as mirroring, rotation, and lung lobe occlusion. Notably, the CT-Net + 3D Convs encoder model achieved a commendable classification accuracy of 0.97 on the independent test set when confronted with images featuring a combination of all three artificial abnormalities. Despite good performance on the surrogate classification tasks, the encoder models struggled to recognize the clinically relevant abnormalities. None of the investigated encoder models managed to decrease the validation loss for the two clinically relevant task: pulmonary nodule and pleural effusion detection. The results also demonstrate that the proposed approach is able to generate accurate and coherent natural language descriptions of the artificial abnormalities. Notably, the Transformer Scratch model achieved a next-word prediction accuracy of 0.84 and a factual accuracy of 0.65 on the independent test set when trained to describe images featuring a combination of all artificial abnormalities. Unfortunately, no clinical reports were generated due to the inability of the encoder models to correctly encode the images with clinically relevant abnormalities.

\textbf{The findings presented in this work confirm that CNN-LLM encoder-decoder models can accurately recognize artificial abnormalities in volumetric CT images and fairy accurately describe these findings using natural language to form narrative-style reports.} The success observed regarding the surrogate tasks aligns with the hypothesis, affirming that a proof of concept is achievable given the current state of CNN and LLM technology. However, the lack of success in detecting clinically relevant abnormalities highlights the challenges posed by the complexity of real-world abnormalities and underscores the need for additional resources such as high quality data and computational power to address these challenges effectively \cite{Goodfellow2016, Kaplan2020}. 

The value of this work lies not in its immediate findings. After all, there is no practical benefit to solving the surrogate tasks. The real value of this work extends beyond its immediate findings, serving as proof of concept and inspiration for further research in this domain. Additionally, this work offers a tested methodology and a publicly available PyTorch Lightning framework, providing a robust foundation for future investigations (Github: \href{https://github.com/Marijn311/CT-Report-Generation}{https://github.com/Marijn311/CT-Report-Generation}, also see Appendices \ref{appendices:framework} "Benefits of the Proposed Framework").

The CT-Net + 3D Convs model emerged as the top-performing encoder model in this study. This superior performance may be attributed to the unique capability of the 3D Convs classifier to model non-linear relationships in the feature maps across all three dimensions, unlike the other classifiers that rely on weighted averaging of features. The significance of this distinction is likely further emphasized by the fact that the convolutions in the feature extractor were not fine-tuned in this study. This explanation is supported further by the comparable results observed in the Med-Net + 3D Convs model. 

The Transformer Scratch model stood out as the most effective decoder model, based on factual accuracy. This finding suggests that a compact model trained from scratch (with 4 million parameters) may be better suited for highly specific tasks with limited variation in target reports, as is case in this work, compared to larger pre-trained models such as Bio+Clinical BERT (with 159 million parameters) and BioGPT (with 346 million parameters). Another noteworthy observation is the subpar performance of the BioGPT model in terms of factual accuracy. This deficiency indicates that the BioGPT model struggles to properly attend to the image content. One plausible explanation could be the depth of the model, with 24 transformer blocks stacked on top of each other, the BioGPT model is the deepest model investigated. Additionally, in the BioGPT model, images are only fed into the model as the first part of the target report. In contrast, both the Transformer Scratch and Bio+Clinical BERT models employ cross-attention mechanisms to incorporate image information at each transformer block, likely enhancing their ability to attend to image details throughout the decoding process.

In the results presented in Section \ref{sec:results} and summarized in Table \ref{tab:surrogate_encoder_accuracy}, it is evident that not all encoder models achieved a classification performance surpassing random guessing across all surrogate tasks. This phenomena may be due to the sensitivity of the encoder models to hyperparameter configuration. Notably, the models which failed to surpass random guessing on the validation set, also failed to substantially diminish training loss. Therefore, it is conceivable that these models could attain success in all tasks with more efforts directed towards identifying successful hyperparameter configurations.

In the results presented in Section \ref{sec:results} and summarized in Table \ref{tab:surrogate_decoder_factual_accuracy}, a notable drop in factual accuracy was observed when transitioning from the individual surrogate tasks to the combined surrogate task. This effect may be related to the number of tokens used to represent the encoded images in the decoder models. In the current study, 100 tokens were arbitrarily chosen to represent the encoded images. For the individual surrogate tasks of rotation and lobe occlusion, there are a maximum of five possible states which the images could have (only two states for the mirroredness task). However, when combining the three surrogate tasks, there were 2*5*5 = 50 possible combinations. Thus, it is plausible that while 100 image tokens were sufficient to capture five different states, they may be inadequate to accurately capture 50 different states. Another possible explanation is that the dataset size was too small for the number of classes. For the rotation and lobe occlusion tasks, there were five output classes each (one class for the mirroredness task). In contrast, for the combined surrogate task, there were 1+5+5 = 11 classes. This means that when moving from the individual to the combined tasks, the number of classes more than doubled while the number of samples in the training dataset remained constant at 1820. This discrepancy could have contributed to the degradation in factual accuracy observed for the combined surrogate task.

An interesting discovery was made regarding the representation of the encoded images (refer to Appendices \ref{appendices:representation} for detailed technical information regarding the different representations). For practical reasons, the encoded images used in this study were highly condensed and generated by fine-tuned classifiers. However, as a final test, the best-performing decoder model was also trained using a less condensed version of the encoded images which were generated by CNNs whose weights were never fine-tuned. Interestingly, the approach using the unoptimized, less condensed representation significantly outperformed the fine-tuned condensed format in terms of the factual accuracy of the generated reports (see Section \ref{sec:results}, Table \ref{tab:surrogate_decoder_factual_accuracy}). This finding confirms the hypothesis that less condensed CNN feature map representations are preferred over highly condensed classifier representations due to their richer information content. However, this result disproves the assumption that unoptimized feature map representation lacks critical information; or, at least, it suggests that the benefit of richer encoded images outweighs the need for optimization of the encoded images for generating accurate reports. These unexpected outcomes highlight the need for further research with more computational resources to explore the impact of providing decoders with fine-tuned less condensed encoded images.

Despite the promising results achieved on surrogate tasks, the study encountered limitations in addressing clinically relevant abnormalities. These limitations can likely be attributed to several factors, including a lack of high-quality data and computational resources \cite{Goodfellow2016, Kaplan2020}. 

The dataset, which includes 464 images (285 with pulmonary nodules and 37 with pleural effusion), is likely insufficient for effective training of classification models. The only comparable study available \cite{Hamamci2024} suggests that larger datasets, exceeding fifty-fold the size utilized here, is necessary for successfully solving clinical tasks.

Another contributing factor to this challenge likely stems from the quality of both the labels and reports within the dataset. The available raw radiology reports often lacked completeness, occasionally serving merely as an addendum or addressing specific questions posed by requesting physicians. Consequently, not all relevant information from the images finds its way into the reference reports. Additionally, the SARLE tool, utilized for mining classification labels from narrative-style reports, introduces a margin of error. As demonstrated in Appendices \ref{appendices:SARLE} Table \ref{tab:sarle-metrics}, the tool exhibited 92\% accuracy in a small-scale test for pulmonary nodule label extraction. It is reasonable to infer that the collective impact of SARLE's inaccuracies and the occasional incompleteness of radiology reports has adversely affected model training for the clinical tasks.

Another notable constraint pertains to the computational resources available for the study. The large size of volumetric data necessitated image downsizing, resulting in a loss of resolution that complicates the visual detection of small features such as sub-centimeter nodules and small effusions. Even after this downsizing, the feature extractor models could only handle one image at a time due to video memory constraints. Operating with a batch size of 1 is likely inadequate for ensuring stable training dynamics. Although gradient accumulation techniques could partially alleviate this issue, they slowed down the training process, requiring hours for a mere 100 gradient update steps. This pace proved incompatible with this work's time constraints, leading to the decision to not fine-tune the feature extractor weights. Consequently, the classifiers were trained on unoptimized image features, which potentially lack critical information regarding the presence or absence of visually small abnormalities. Conversely, the artificial abnormalities such as mirroring, rotation, and lung occlusion affect numerous pixels or all pixels in the image, making it more likely that the unoptimized features capture these abnormalities. This difference may explain why the surrogate tasks exhibited high accuracy while classification of clinically relevant abnormalities consistently failed to surpass random guessing.

Moving forward, several avenues for future research naturally emerge from the results and limitations of this work. Future efforts could be focused on attaining more high-quality data and computational resources. With increased resources, subsequent research endeavors could readdress the clinically relevant tasks by 1) using a larger dataset, 2) fine-tuning the feature extractor, and 3) using a large latent space to provide the decoder with richer features, see Appendices \ref{appendices:representation}. Additionally, the encoder-decoder could be trained end-to-end, allowing the model itself to learn which features should be extracted for generating the most accurate radiology reports, instead of manually defining the classification tasks. Simultaneously, an end-to-end approach takes away the inaccuracies introduced by the SARLE tool.  

Another potentially interesting expansion, involves incorporating location data alongside abnormality labels. While the original SARLE tool \cite{Draelos2021} offered the capability to mine location information for each abnormality, this feature was omitted in the present work for simplicity. Nevertheless, radiology reports frequently delineate both the abnormality and its location. By training the encoder to classify both the abnormality and its location, the encoded images may provide richer information for the decoder.

An additional avenue for extending this research involves investigating the use of auxiliary tasks or custom loss functions to enhance the factual accuracy of the generated reports. Presently, factual accuracy appears to be the primary limitation preventing these models from achieving practical relevance.

Yet another way to potentially expand this work involves shifting focus to other domains. The methodology and framework developed in this work holds relevance beyond the confines of radiology, potentially finding applications in fields such as aiding the visually impaired by describing what can be seen. As another example, this framework may also assist in automatic compressing of surveillance footage by generating concise reports detailing the time and location of abnormal occurrences, thereby eliminating the need for exhaustive video storage and manual examination.

In conclusion, this study aimed to develop a proof of concept model capable of automatically identifying abnormalities in volumetric CT images and describing them using natural language. Results indicate promising performance in classifying CT images with synthetic abnormalities, but struggles persist in recognizing clinically relevant abnormalities. The findings validate the feasibility of using the proposed encoder-decoder approach for this task, while highlighting the need for more data and computational resources. Despite limitations, this work offers a working proof of concept, tested methodology, and coding framework for future research.

\bibliographystyle{unsrt}  
\bibliography{references}  

\begin{thebibliography}{10}

\bibitem{rcr-census}
{The Royal College of Radiologists}.
\newblock {Clinical Radiology Workforce Census}.
\newblock {\em Available at: https://www.rcr.ac.uk/news-policy/policy-reports-initiatives/clinical-radiology-census-reports/}, 2022.

\bibitem{Konstantinidis2023}
Kleanthis Konstantinidis.
\newblock {The shortage of radiographers: A global crisis in healthcare.}
\newblock {\em Journal of medical imaging and radiation sciences}, 10 2023.

\bibitem{IHS2021}
{IHS Markit Ltd.}
\newblock {\em {The Complexities of Physician Supply and Demand: Projections From 2019-2034}}.
\newblock 2021.

\bibitem{Farmakis2021}
Shannon~G Farmakis, Jocelyn~D Chertoff, and Richard~A Barth.
\newblock {Pediatric Radiologist Workforce Shortage: Action Steps to Resolve.}
\newblock {\em Journal of the American College of Radiology : JACR}, 18(12):1675--1677, 12 2021.

\bibitem{DeBenedectis2022}
Carolynn~M DeBenedectis, Lucy~B Spalluto, Lisa Americo, Casey Bishop, Asim Mian, David Sarkany, Nolan~J Kagetsu, and Priscilla~J Slanetz.
\newblock {Health Care Disparities in Radiology-A Review of the Current Literature.}
\newblock {\em Journal of the American College of Radiology : JACR}, 19(1 Pt B):101--111, 1 2022.

\bibitem{Shoshan2022}
Yoel Shoshan, Ran Bakalo, Flora Gilboa-Solomon, Vadim Ratner, Ella Barkan, Michal Ozery-Flato, Mika Amit, Daniel Khapun, Emily~B. Ambinder, Eniola~T. Oluyemi, Babita Panigrahi, Philip~A. DiCarlo, Michal Rosen-Zvi, and Lisa~A. Mullen.
\newblock {Artificial Intelligence for Reducing Workload in Breast Cancer Screening with Digital Breast Tomosynthesis}.
\newblock {\em Radiology}, 303(1):69--77, 2022.

\bibitem{Tong2023}
Wen-Juan Tong, Shao-Hong Wu, Mei-Qing Cheng, Hui Huang, Jin-Yu Liang, Chao-Qun Li, Huan-Ling Guo, Dan-Ni He, Yi-Hao Liu, Han Xiao, Hang-Tong Hu, Si-Min Ruan, Ming-De Li, Ming-De Lu, and Wei Wang.
\newblock {Integration of Artificial Intelligence Decision Aids to Reduce Workload and Enhance Efficiency in Thyroid Nodule Management.}
\newblock {\em JAMA network open}, 6(5):e2313674, 5 2023.

\bibitem{Sim2023}
Jordan Z~T Sim, K~N Bhanu~Prakash, Wei~Min Huang, and Cher~Heng Tan.
\newblock {Harnessing artificial intelligence in radiology to augment population health.}
\newblock {\em Frontiers in medical technology}, 5:1281500, 2023.

\bibitem{Zhao2022}
Kai Zhao, Shuai Ma, Zhaonan Sun, Xiang Liu, Ying Zhu, Yufeng Xu, and Xiaoying Wang.
\newblock {Effect of AI-assisted software on inter- and intra-observer variability for the X-ray bone age assessment of preschool children}.
\newblock {\em BMC Pediatrics}, 22(1):1--6, 2022.

\bibitem{Krizhevsky2012}
Alex Krizhevsky, Ilya Sutskever, and Geoffrey~E Hinton.
\newblock {ImageNet classification with deep convolutional neural networks}.
\newblock In {\em Proceedings of the 25th International Conference on Neural Information Processing Systems - Volume 1}, NIPS'12, page 1097–1105, Red Hook, NY, USA, 2012. Curran Associates Inc.

\bibitem{Hochreiter1997}
Sepp Hochreiter and Jürgen Schmidhuber.
\newblock {Long Short-Term Memory}.
\newblock {\em Neural Computation}, 9(8):1735--1780, 1997.

\bibitem{Vaswani2017}
Ashish Vaswani, Noam Shazeer, Niki Parmar, Jakob Uszkoreit, Llion Jones, Aidan~N Gomez, Łukasz Kaiser, and Illia Polosukhin.
\newblock {Attention is All you Need}.
\newblock In I~Guyon, U~Von Luxburg, S~Bengio, H~Wallach, R~Fergus, S~Vishwanathan, and R~Garnett, editors, {\em Advances in Neural Information Processing Systems}, volume~30. Curran Associates, Inc., 2017.

\bibitem{Devlin2019}
Jacob Devlin, Ming-Wei Chang, Kenton Lee, and Kristina Toutanova.
\newblock {BERT: Pre-training of Deep Bidirectional Transformers for Language Understanding}.
\newblock In Jill Burstein, Christy Doran, and Thamar Solorio, editors, {\em Proceedings of the 2019 Conference of the North American Chapter of the Association for Computational Linguistics: Human Language Technologies, Volume 1 (Long and Short Papers)}, pages 4171--4186, Minneapolis, Minnesota, 6 2019. Association for Computational Linguistics.

\bibitem{Radford2018}
Alec Radford, Karthik Narasimhan, Tim Salimans, and Ilya Sutskever.
\newblock {Improving language understanding by generative pre-training}.
\newblock 2018.

\bibitem{Yin2019}
Changchang Yin, Buyue Qian, Jishang Wei, Xiaoyu Li, Xianli Zhang, Yang Li, and Qinghua Zheng.
\newblock {Automatic generation of medical imaging diagnostic report with hierarchical recurrent neural network}.
\newblock {\em Proceedings - IEEE International Conference on Data Mining, ICDM}, 2019-Novem(Icdm):728--737, 2019.

\bibitem{Kaur2022}
Navdeep Kaur and Ajay Mittal.
\newblock {RadioBERT: A deep learning-based system for medical report generation from chest X-ray images using contextual embeddings.}
\newblock {\em Journal of biomedical informatics}, 135:104220, 11 2022.

\bibitem{Nimalsiri2023}
Wimukthi Nimalsiri, Mahela Hennayake, Kasun Rathnayake, Thanuja~D. Ambegoda, and Dulani Meedeniya.
\newblock {Automated Radiology Report Generation Using Transformers}.
\newblock {\em ICARC 2023 - 3rd International Conference on Advanced Research in Computing: Digital Transformation for Sustainable Development}, pages 90--95, 2023.

\bibitem{DallaSerra2024}
Francesco Dalla~Serra, Chaoyang Wang, Fani Deligianni, Jeffrey Dalton, and Alison~Q. O’Neil.
\newblock {Finding-Aware Anatomical Tokens for Chest X-Ray Automated Reporting}.
\newblock {\em Lecture Notes in Computer Science (including subseries Lecture Notes in Artificial Intelligence and Lecture Notes in Bioinformatics)}, 14348 LNCS:413--423, 2024.

\bibitem{DallaSerra2023}
Francesco Dalla~Serra, Chaoyang Wang, Fani Deligianni, Jeffrey Dalton, and Alison~Q O'Neil.
\newblock {Controllable Chest X-Ray Report Generation from Longitudinal Representations}.
\newblock 10 2023.

\bibitem{Tanida2023}
T~Tanida, P~Muller, G~Kaissis, and D~Rueckert.
\newblock {Interactive and Explainable Region-guided Radiology Report Generation}.
\newblock In {\em 2023 IEEE/CVF Conference on Computer Vision and Pattern Recognition (CVPR)}, pages 7433--7442, Los Alamitos, CA, USA, 6 2023. IEEE Computer Society.

\bibitem{Xue2024}
Youyuan Xue, Yun Tan, Ling Tan, Jiaohua Qin, and Xuyu Xiang.
\newblock {Generating radiology reports via auxiliary signal guidance and a memory-driven network}.
\newblock {\em Expert Systems with Applications}, 237:121260, 2024.

\bibitem{Chen2020}
Zhihong Chen, Yan Song, Tsung-Hui Chang, and Xiang Wan.
\newblock {Generating Radiology Reports via Memory-driven Transformer}.
\newblock 10 2020.

\bibitem{Jing2017}
Baoyu Jing, Pengtao Xie, and Eric Xing.
\newblock {On the Automatic Generation of Medical Imaging Reports}.
\newblock 11 2017.

\bibitem{Liao2023}
Yuxiang Liao, Hantao Liu, and Irena Spasi{\'{c}}.
\newblock {Deep learning approaches to automatic radiology report generation: A systematic review}.
\newblock {\em Informatics in Medicine Unlocked}, 39:101273, 2023.

\bibitem{Monshi2020}
Maram Mahmoud~A Monshi, Josiah Poon, and Vera Chung.
\newblock {Deep learning in generating radiology reports: A survey.}
\newblock {\em Artificial intelligence in medicine}, 106:101878, 6 2020.

\bibitem{Dai2022}
Yin Dai, Yifan Gao, Fayu Liu, and Jun Fu.
\newblock {Mutual Attention-based Hybrid Dimensional Network for Multimodal Imaging Computer-aided Diagnosis}.
\newblock 2022.

\bibitem{Zhou2019}
Zongwei Zhou, Vatsal Sodha, Jiaxuan Pang, Michael~B Gotway, and Jianming Liang.
\newblock {Models Genesis}.
\newblock {\em Medical Image Analysis}, 67:101840, 2021.

\bibitem{Draelos2021}
Rachel~Lea Draelos, David Dov, Maciej~A Mazurowski, Joseph~Y Lo, Ricardo Henao, Geoffrey~D Rubin, and Lawrence Carin.
\newblock {Machine-learning-based multiple abnormality prediction with large-scale chest computed tomography volumes}.
\newblock {\em Medical Image Analysis}, 67:101857, 2021.

\bibitem{Ray2023}
Partha~Pratim Ray.
\newblock {ChatGPT: A comprehensive review on background, applications, key challenges, bias, ethics, limitations and future scope}.
\newblock {\em Internet of Things and Cyber-Physical Systems}, 3(March):121--154, 2023.

\bibitem{Vinyals2015}
O~Vinyals, A~Toshev, S~Bengio, and D~Erhan.
\newblock {Show and tell: A neural image caption generator}.
\newblock In {\em 2015 IEEE Conference on Computer Vision and Pattern Recognition (CVPR)}, pages 3156--3164, Los Alamitos, CA, USA, 6 2015. IEEE Computer Society.

\bibitem{Russakovsky2015}
Olga Russakovsky, Jia Deng, Hao Su, Jonathan Krause, Sanjeev Satheesh, Sean Ma, Zhiheng Huang, Andrej Karpathy, Aditya Khosla, Michael Bernstein, Alexander~C Berg, and Li~Fei-Fei.
\newblock {ImageNet Large Scale Visual Recognition Challenge}.
\newblock {\em Int. J. Comput. Vision}, 115(3):211–252, 12 2015.

\bibitem{Zeller2014}
Matthew~D Zeiler and Rob Fergus.
\newblock {Visualizing and Understanding Convolutional Networks}.
\newblock In David Fleet, Tomas Pajdla, Bernt Schiele, and Tinne Tuytelaars, editors, {\em Computer Vision -- ECCV 2014}, pages 818--833, Cham, 2014. Springer International Publishing.

\bibitem{Goodfellow2016}
Ian Goodfellow, Yoshua Bengio, and Aaron Courville.
\newblock {\em {Deep Learning}}.
\newblock MIT Press, 2016.

\bibitem{Jiang2018}
Tian Jiang, Li~Cong, Shi Zhongchao, and Xu~Feiyu.
\newblock {A Diagnostic Report Generator from CT Volumes on Liver Tumor with Semi-supervised Attention Mechanism}.
\newblock pages 517--524, 2018.

\bibitem{Hamamci2024}
Ibrahim~Ethem Hamamci, Sezgin Er, and Bjoern Menze.
\newblock {CT2Rep: Automated Radiology Report Generation for 3D Medical Imaging}.
\newblock 2024.

\bibitem{Ye2018}
Qing Ye, Jing Wu, Yihuan Lu, Mika Naganawa, Jean-Dominique Gallezot, Tianyu Ma, Yaqiang Liu, Lynn Tanoue, Frank Detterbeck, Justin Blasberg, Ming-Kai Chen, Michael Casey, Richard~E Carson, and Chi Liu.
\newblock {Improved discrimination between benign and malignant LDCT screening-detected lung nodules with dynamic over static (18)F-FDG PET as a function of injected dose.}
\newblock {\em Physics in medicine and biology}, 63(17):175015, 9 2018.

\bibitem{Wasserthal2023}
Jakob Wasserthal, Hanns-Christian Breit, Manfred~T Meyer, Maurice Pradella, Daniel Hinck, Alexander~W Sauter, Tobias Heye, Daniel~T Boll, Joshy Cyriac, Shan Yang, Michael Bach, and Martin Segeroth.
\newblock {TotalSegmentator: Robust Segmentation of 104 Anatomic Structures in CT Images.}
\newblock {\em Radiology. Artificial intelligence}, 5(5):e230024, 9 2023.

\bibitem{Chen2019}
Sihong Chen, Kai Ma, and Yefeng Zheng.
\newblock {Med3D: Transfer Learning for 3D Medical Image Analysis}.
\newblock 2019.

\bibitem{Alsentzer2019}
Emily Alsentzer, John~R. Murphy, Willie Boag, Wei-Hung Weng, Di~Jin, Tristan Naumann, and Matthew B.~A. McDermott.
\newblock {Publicly Available Clinical BERT Embeddings}.
\newblock 4 2019.

\bibitem{Luo2022}
Renqian Luo, Liai Sun, Yingce Xia, Tao Qin, Sheng Zhang, Hoifung Poon, and Tie-Yan Liu.
\newblock {BioGPT: generative pre-trained transformer for biomedical text generation and mining}.
\newblock {\em Briefings in Bioinformatics}, 23(6), 2022.

\bibitem{Kaplan2020}
Jared Kaplan, Sam McCandlish, Tom Henighan, Tom~B Brown, Benjamin Chess, Rewon Child, Scott Gray, Alec Radford, Jeffrey Wu, and Dario Amodei.
\newblock {Scaling laws for neural language models}.
\newblock {\em arXiv preprint arXiv:2001.08361}, 2020.

\end{thebibliography}

\section{Appendices} \label{sec:appendices}
    \subsection{SARLE Evaluation}\label{appendices:SARLE}
To validate SARLE's label extraction performance, a small manually labeled dataset of 25 reports (123 sentences) was utilized. These reports were sourced from an independent hospital (Antoni van Leeuwenhoek hospital) that was not included in the original dataset. The evaluation was only done using the pulmonary nodule labels. Evaluation using the pleural effusion was omitted due to its low prevalence. Evaluation results, presented in Tables \ref{tab:confusion-matrix-sarle} and \ref{tab:sarle-metrics}, revealed 9 true negatives, 14 true positives, 0 false positives, and 2 false negatives, yielding an accuracy of 0.92, sensitivity of 1.0, and specificity of 0.875. While acknowledging the limited sample size, these results suggest sufficient accuracy in automatic label extraction to justify trying model training on these labels.

\begin{table}[h]
\centering
\caption{Confusion Matrix for SARLE's "Pulmonary Nodule" Label Extraction on Independent Test Set (N=25)}
\label{tab:confusion-matrix-sarle}
\begin{tabular}{cccc}
\multicolumn{2}{c}{\textbf{}} & \multicolumn{2}{c}{\textbf{Prediction}} \\
\cmidrule{3-4}
\multicolumn{2}{c}{\textbf{}} & \textbf{Positive} & \textbf{Negative} \\
\midrule
\multirow{2}{*}{\textbf{Ground Truth}} & \textbf{Positive} & 9 & 0 \\
\cmidrule{2-4}
& \textbf{Negative} & 2 & 14 \\
\bottomrule
\end{tabular}
\end{table}

\begin{table}[h]
\centering
\caption{Classification Metrics for SARLE's "Pulmonary Nodule" Label Extraction on Independent Test Set (N=25)}
\label{tab:sarle-metrics}
\begin{tabular}{llcc}
\toprule
\multicolumn{2}{l}{\textbf{Metric}} & \textbf{Formula} & \textbf{Value} \\
\toprule
\multicolumn{2}{l}{Accuracy} & $ \frac{TP + TN}{TP + TN + FP + FN}$ & 0.92\\
\thinnermidrule
\multicolumn{2}{l}{Sensitivity (Recall)} & $\frac{TP}{TP + FN}$ & 1.0\\
\thinnermidrule
\multicolumn{2}{l}{Specificity} & $\frac{TN}{TN + FP}$ & 0.875\\
\thinnermidrule
\end{tabular}
\end{table}

\subsection{Token Representation Versus Feature Map Representation}\label{appendices:representation}
The developed framework allows the encoded images to be passed to the decoder in two different formats: the "token representation" and the "feature map representation".

To ensure technical compatibility with the decoder models, the encoded images must conform to the specific shape expected by these models. The decoder models were originally designed to process text, thus they expect an input shape of [batch\_size, nr\_tokens, embedding\_vector]. However, the encoder's CT-Net feature extractor produces tensors with the shape [batch\_size, nr\_3\_slice\_chunks, nr\_channels, feature\_map\_dimension\_1, feature\_map\_dimension\_2]. This tensor format is defined as the feature map representation.

When using the feature map representation for the decoder models, the nr\_channels dimension is reduced via mean averaging to decrease the tensor size. Next, the two feature map dimensions are concatenated to form a single feature vector, which is then reshaped into the embedding\_vector dimension using a trainable linear layer. The nr\_3\_slice\_chunks dimension becomes the nr\_tokens dimension without further processing.

It is important to note that in this study, the feature extractors were not fine-tuned. This implies that the feature maps may lack the necessary information for generating factually accurate reports. To address this issue, various classifiers were fine-tuned on the feature maps, and the encoded images were obtained by extracting the tensor from the penultimate dense layer of the classifier. This tensor has the shape [batch\_size, encoded\_image\_vector]. This representation is termed the "token representation" because the floating-point values in the encoded\_image\_vector are transformed into integers within the range of 0-99 through linear rescaling and rounding operations. The resulting integers, representing image tokens, have thus become the required nr\_tokens dimension. These tokens are subsequently embedded using an embedding layer to form the necessary embedding\_vector dimension.

Due to limited computational resources and the assumption that unoptimized feature maps might lack critical information, token representation was used throughout this study. However, for the selected decoder model tested on an independent test set containing a combination of artificial abnormalities, both the token representation and the feature map representation were evaluated.

Interestingly, the approach using the unoptimized feature map representation significantly outperformed the token representation in terms of the factual accuracy of the generated reports in Table \ref{tab:surrogate_decoder_factual_accuracy}. This finding confirms the hypothesis that the feature map representation is favored because it contains richer information than the token representation. On the other hand, this result disproves the assumption that unoptimized feature maps may lack (too much) critical information; or, at least, it suggests that the benefit of richer encoded images outweighs the need for optimization of the feature maps for generating accurate reports. These unexpected outcomes highlight the need for further research with more computational resources to explore the impact of providing decoders with fine-tuned feature maps.

\subsection{Benefits of the Proposed Framework}\label{appendices:framework}
The developed framework is designed for ease of adaptation by researchers with varying resources. Notably, it offers user-friendly flexibility in scaling computational resources, as the number of GPUs and data parallelization techniques can be adjusted by a single parameter in the configuration file. Moreover, the provided framework also offers user-friendly flexibility in scaling the dataset, allowing for seamless integration of additional data at any point without the need for extensive re-processing. Furthermore, the framework provides the image pre-processing pipeline and the classification label mining pipeline, thereby taking away most of the manually work needed to clean and annotate new data. Lastly, the framework's modular structure allows for seamless addition of extra model architecture. 

\end{document}